\documentclass[manuscript]{aastex}

\usepackage{rotating}

\shorttitle{}

\shortauthors{Yang et al.}

\begin{document}

\title{Global explicit particle-in-cell simulations of the nonstationary bow shock and magnetosphere}

\author{Zhongwei Yang\altaffilmark{1}, Can Huang\altaffilmark{2}, Ying D. Liu\altaffilmark{1}, George K. Parks \altaffilmark{3}, Rui Wang \altaffilmark{1}, Quanming Lu\altaffilmark{2}, Huidong Hu \altaffilmark{1}}

\altaffiltext{1}{State Key Laboratory of Space Weather, National Space Science Center, Chinese Academy of Sciences, Beijing 100190, China;
zwyang@spaceweather.ac.cn}

\altaffiltext{2}{CAS Key Laboratory of Geospace Environment, Department of Geophysics and Planetary Science, University of Science and Technology of China, Hefei, China}

\altaffiltext{3}{Space Sciences Laboratory, University of California, Berkeley, USA}

\begin{abstract}

We carry out \textbf{two dimensional (2D)} global Particle-in-Cell (PIC) simulations of the interaction between the solar wind and a dipole field to study the formation of the bow shock and magnetosphere. A self-reforming bow shock ahead of a dipole field is presented by using relatively high temporal-spatial resolutions. We find that (1) the bow shock and the magnetosphere are formed and reach a quasi-stable state after \textbf{several} ion cyclotron periods, (2) under the $B_z$ southward solar wind condition, the bow shock undergoes the self-reformation for low $\beta_i$ and high $M_A$. Simultaneously, a magnetic reconnection in the magnetotail is found. For high $\beta_i$ and low $M_A$, \textbf{the} shock becomes quasi-stationary, and the magnetotail reconnection disappears, (3) The magnetopause deflects the magnetosheath plasmas. The sheath particles injected at the quasi-perpendicular region of the bow shock can be convected to downstream of an oblique shock region. A fraction of these sheath particles can leak out from the magnetosheath at the wings of the bow shock. Hence the downstream situation is more complicated than that for a planar shock produced in local simulations.

\end{abstract}


\section{Introduction}

Collisionless shocks are of particular interest for space, plasma and astrophysics. In the shock transition the bulk energy of the plasma is converted into thermal energy in the absence of particle collisions \citep{Tidmann1971,Lembege2004,Burgess2005}. The collisionless shock has been studied for many decades. However, the cyclic reformation of the shock structure is still a major unresolved issue for collisionless shock physics \citep{Winske1990,Scholer2003,Hada2003,Lee2004,Burgess2007,Umeda2008,Yang2009,Rekaa2014}. The term ¡°self-reformation¡± describes a process where the particles reflected by the shock ramp accumulate ahead of the shock and form a shock foot, which then grows and becomes a new ramp. The new ramp starts to reflect incident particles, and the process repeats itself. One striking point is that the ramp width can be very narrow covering a few electron inertial lengths during the reformation cycle \citep{Scholer2003,Hada2003,Mazelle2010}. At such narrow shock ramps, the cross shock electric field is large and the incident particles can be efficiently accelerated to very high energies \citep{Lee1996,Zank1996,Yang2009}. The shock front self-reformation was initially predicted from one dimensional hybrid \citep{Quest1985} and Particle-in-Cell (PIC) \citep{Lembege1987} simulations. This problem has been also investigated by theoretical studies \citep{Krasnoselskikh2002}. In situ measurements of the terrestrial bow shock made by the CLUSTER mission \citep{Moullard2006,Lobzin2007,Mazelle2010} clearly show the shock front is strongly nonstationary. The shock front nonstationarity is also important for the bow shocks at other planets (for example, \textbf{Uranus and Mercury}) and the heliospheric termination shock \citep{Burlaga2008,Richardson2008,Tiu2011,Sundberg2013}. Thus the shock front nonstationarity is a widespread natural physical phenomenon. However, the effects of curved geometry and the solar wind condition on the bow shock front nonstationarity are still not well understood.

Although global magnetohydrodynamics (MHD) simulations can successfully describe the macrostructures \citep{Ogino1986,Wang2014} and macroinstabilities \citep{Farrugia1998,Li2013} of the magnetosphere, the hybrid and PIC simulations are needed to solve the problems of microstructures on ion and electron scales and microinstabilities. \textbf{During the past two decades, global hybrid simulations have been performed to study bow shocks of different planets \citep{Swift1995,Lin2003,Omidi2005,Omidi2006,Travnicek2007,Lu2015}. \citet{Turc2015} studied the interaction of a magnetic cloud with a bow shock by using a global hybrid simulation with a spatial resolution of 1 ion inertial length. Studies on foreshock and hot flow anomalies also requires at least hybrid simulations \citep{Omidi2007,Omidi2010,Omidi2014} instead of MHD models. A new hybrid-Vlasov code has been developed to investigate the ion distributions in the Earth's foreshock and to explain the THEMIS observations \citep{Kempf2015,Palmroth2015}. The hybrid-Vlasov approach ensures a uniform sampling of the ion distribution function in all spatial and velocity dimensions, as the full three-dimensional ion velocity distribution function is propagated in each real space cell. However, it has a very high computational cost. In all of the hybrid models above, an artificial/anomalous resistivity must be included to generate dissipation on electron scales.} \citet{Lipatov1999} find that the different values of the resistivity can lead to very different shock structures.

\textbf{In the global PIC simulation \citep{Buneman1995,Cai2015,Peng2015a}, which has higher numerical noise but lower computational costs than the hybrid-Vlasov simulations \citep{Kempf2015,Palmroth2015}, the accessibility to both ion and electron scales is automatically and self-consistently included.} \citet{Buneman1995} created a three dimensional electromagnetic full PIC code (TRISTAN) to simulate the solar wind-magnetosphere interaction. In recent years, \citet{Cai2015} \textbf{parallelized} the TRISTAN code and obtained a global structure of the magnetosphere with a resolution of $\sim$ 0.1 ion inertial length. This setup however has difficulty to retrieve the nonstationary shock front as in previous local simulations \citep{Lembege1987,Hada2003,Yang2009}. \citet{Peng2015a,Peng2015b} recently improved the three dimensional implicit PIC code (iPic3D) using a dipole magnetic field immersed in the flow of the plasma which showed the formation of a magnetosphere. The highest spatial and temporal resolutions used in their work are 0.05$d_i$ and 0.15$\omega_{pi}^{-1}$, where $d_i$ and $\omega_{pi}^{-1}$ are the ion inertial length and the ion plasma frequency respectively. In their simulation with the highest resolutions, the total size of the simulation box is 20$d_i\times$20$d_i\times$20$d_i$, and the dipole becomes very small. Therefore, the resulting bow shock is blurred and is on a scale of several ion inertial lengths, which is roughly the same as the shock front rippling scale observed in local PIC simulations \citep{Hellinger2007,Lembege2009}. Hence, it is difficult to retrieve the features of shock front nonstationarity. However, 3D global simulations with \textbf{high performance configuration} (high resolutions, large particle number per cell, large simulation domain etc.) are still a very computation consuming task. It is thought that the details of the bow shock and the magnetosphere can eventually be well obtained from the iPic3D code \citep{Peng2015a,Peng2015b}, and we expect to see the fascinating results in the near future.

In this paper, we use a two dimensional (2D) explicit full PIC code to achieve a relatively high spatial and temporal resolution for simulating the nonstationary bow shock. A self-reforming curved bow shock is generated by the interaction of the solar wind with a 2D dipole field. The paper is organized as follows. In section 2, we describe briefly the simulation model. The results from our global simulations are presented in section 3 followed by a discussion of the results and conclusions in section 4.

\section{Simulation model}

\textbf{Previous high resolution hybrid simulations ($\Delta x=0.1d_i$) have shown that the artificial/anomalous resistivity $\eta$ employed in hybrid simulations can strongly affect the shock front self-reformation process in the same plasma and Mach number conditions ($M_A\sim2.1$). \citet{Lembege2009} find that: (1) for the high resistivity run ($\eta=10^{-2}\mu_0v_A^2\omega_{ci}^{-1}$), the steepening of the ramp is restricted and cannot initiate a self-reformation even if the spatial resolution is relatively high, and (2) for the low resistivity run ($\eta=10^{-4}\mu_0v_A^2\omega_{ci}^{-1}$), the self-reformation is partial in the sense that the foot amplitude is increasing but stays relatively weak and never reaches an amplitude comparable to that of the old ramp. Instead, the new ramp crashes down and restarts reflecting new incoming ions. They conclude that using a high resistivity value and a low spatial resolution will stop almost simultaneously the self-reformation and the emission of nonlinear whistler waves. But for very high Mach number shocks ($M_A\sim23$), the self-reformation can appear in hybrid simulations with a high resistivity value ($\eta=10^{-2}\mu_0v_A^2\omega_{ci}^{-1}$) and a relatively high spatial resolution \citep{Tiu2011}. Hence, the chosen of different resistivity can lead to different shock front structures as mentioned by \citet{Lipatov1999}. In fact, it is difficult to tell how much the artificial resistivity we should choose. In addition, the use of high resolution hybrid simulations can be questionable in terms of computing costs when compared to 1D/2D PIC simulations performed with a reasonable mass ratio. Let us note that the question of accessibility to a small scale (lower than ion scale) is expressed differently in full PIC and hybrid simulations. In PIC simulations, the accessibility to both ion and electron scales is automatically and self-consistently included. Furthermore, the calculations of the electric field in hybrid and full PIC codes are different. The former usually use the generalized Ohm's law with an artificial resistivity, and the latter is based on the complete Maxwell equations without an artificial resistivity.}

\textbf{For that reason, a 2D full particle code is used to simulate the bow shock formation due to the solar wind-magnetosphere interaction.} \citet{Yang2015} have already described the simulation code used for modeling of planar shocks in considerable detail, hence only a brief description is given here. The simulation plane corresponds to the $x$-$z$ plane with $x$ along the solar wind flow direction (Sun-planet line) and $z$ pointing along the dipole axis (as in GSM coordinate system). Consequently, the $x$-$z$ simulation plane corresponds to the noon-midnight meridian plane with $z$ pointing northward. The solar wind Alfv\'en Mach number ranges from 4 to 8 and the value of $\beta_i$ (the ratio of the plasma pressure to the magnetic pressure) is set from 0.01 to 2. Based on previous local simulation results \citep{Lembege1992,Hada2003,Hellinger2007,Yang2013,Hao2014}, the shock is expected to be in the supercritical regime \citep{Lembege1987} and the shock front can be nonstationary. It is generally believed that the distance between the center of the dipole and the magnetopause at the subsolar point is the shortest when the interplanetary magnetic field (IMF) is shouthward. To save computational time and decrease simulation box, the IMF initially lies in the $x$-$z$ plane, points southward ($-z$), and makes a 90$^{\circ}$ angle with the $x$-axis. A reduced mass ratio $m_i/m_e=20$ is chosen, which is a little larger than in global PIC simulations \citep{Cai2015}. The size of the simulation box is $L_x\times L_z\approx100d_i\times100d_i$, and the grid consists of 4096$\times$4096 cells. For reference, the temporal-spatial resolution used in previous hybrid and PIC simulations is summarized in Table 1. Local hybrid and PIC simulations show that the shock front self-reformation can be seen in high resolution cases, i.e., $\Delta x\le0.05d_i$ \citep{Lembege2009}. The grid spacing used in our simulations is $\Delta x=\Delta z=0.025d_i=0.11d_e=0.68\lambda_{De}$, where $d_e$ and $\lambda_{De}$ are the electron inertial length and the Debye length respectively. The chosen temporal resolution is $\Delta t=0.001\Omega_{ci}^{-1}=0.015\omega_{pi}^{-1}$. The initial setups of the dipole and IMF fields are similar to those used in previous simulations \citep{Omidi2005,Peng2015a}. The particles are evenly distributed from the solar wind to the nightside of the dipole to speed up the magnetosphere formation. In our explicit code, the electric field and the magnetic flux function instead of the magnetic strength are updated at each time step. This will help us directly obtain the accurate magnetic field lines by plotting the contour of the magnetic flux function on the 2D simulation plane.

\section{Simulation results}

We present the overall structure of the bow shock and magnetosphere in the global PIC simulation. Our study is separated into three parts: (1) we first show the time-evolution of the bow shock and the entire magnetosphere; (2) then we show shock front self-reformation at different shock normal angles along the curved shock front, and (3) the impact of the plasma $\beta$ value and upstream solar wind Mach number $M_A$ on the nonstationarity of the bow shock.

\subsection{Formation of the bow shock and magnetosphere}

Figure 1 shows the macroscopic and microscopic evolutions of the normalized ion number density $log_{10}(N_i/N_0+1)$ in the meridional plane (i.e., in the simulation $x-z$ plane). Initially, the magnetosphere undergoes an expanding stage. Figures 1a-1c (right column) show the first stage of a bow shock formation. At $t=0.1\Omega_{ci}^{-1}$ (cycle 100), the high ion density is located at the cusp region and behind the dipole center. At a later time ($t=1-3\Omega_{ci}^{-1}$, cycles 1000-3000), more ions are accumulated in front of the dipole and the cusp region. A fraction of incident ions is reflected at the newborn magnetopause back toward the Sun. Instead of a perfect conductive wall, the solar wind ions are reflected by an elastic wall ``magnetopause". The physical mechanism is similar to that in 1D local simulations of a collisionless shock by the reflection wall method. Figures 1d-1f (center panels) show the second stage of the bow shock formation. At $t=4\Omega_{ci}^{-1}$ (cycle 4000), the magnetotail stretches in the $x$ direction. The plasma sheet and magnetotail lobe are easily distinguished. The plasma mantle (light blue) around the lobe (dark blue) is also stretching accompanied by the plasma sheet. At $t=6\Omega_{ci}^{-1}$ (cycle 6000), the reflected ions ahead of the magnetopause are convected back together with the new incoming solar wind ions, and they are compressed in front of the magnetopause. Then a dense magnetosheath is formed. The mature shock is firstly generated at the subsolar point. At $t=8\Omega_{ci}^{-1}$ (cycle 8000), the bow shock is almost completely formed and the magnetotail continues to stretch. Figures 1g-1i (cycles start from 11000) show that both the bow shock while the entire magnetosphere have now reached a quasi-static state with minimal changes until the end of the simulation. In this low $\beta_i$ (=0.01) and high Mach number ($M_A=8$) case, a nonstationary bow shock front with ion scale ripples of the order of ion inertial length $d_i$ is observed.

An overview of the bulk velocity and electromagnetic field components of the bow shock and the magnetosphere at $t=14\Omega_{ci}^{-1}$ is shown in Figure 2. The magnetic field line (in black) is superimposed on the contours of the ion bulk velocity profile $V_z$ (Figure 2c). In the magnetosheath (downstream region of the bow shock), Figures 2a-2f show that the bow shock reduces the super-Alfv\'enic solar wind speed and the magnetopause deflects the downstream plasma flow. The sheath plasma diverts around the magnetosphere. The bulk velocity components $V_x$ of both ions (Figure 2a) and electrons (Figure 2d) become sub-Alfv\'enic. One striking point is that the velocity component $V_z$ (let alone the total bulk speed) can still remain super-Alfv\'enic in the sheath region due to the deflection motion. The shock only decreases the upstream inflow speed along the shock normal. Furthermore, the velocity moments of ions (Figure 2b) and electrons (Figure 2e) at the magnetopause and the magnetotail plasma sheet are in opposite directions, and they contribute to the magnetopause and cross-tail currents. In the magnetotail, the plasma flow and electromagnetic fields are quite similar to that obtained in local magnetic reconnection simulations \citep{Daughton2006,Fu2006,Lu2010,Huang2014,Guo2015}. Magnetic reconnection takes place in the magnetotail because of the southward magnetic field used in the simulation.

\subsection{A baseline case of a nonstationary bow shock}

To examine the impact of shock normal angle $\theta_{Bn}$ on the shock microstructure and the shock reformation process in our simulation, we show in Figure 3 the time evolution (stack-plots) of the magnetic field components $B_x$, $B_y$, and $-B_z$ for the shock profiles A, B, C, and D measured at different locations (from perpendicular shock to oblique shock regions) of the bow shock. The spatial ranges of measurements are marked by the red lines in Figure 2g. In all panels, the same scale are used for the stack-plots.

Figures 3a-3c show the magnetic field components of the shock profile A, which is a nearly perpendicular shock ($\theta_{Bn}\approx90^o$) measured at $z=51d_i$ in the bow shock (see Figure 2g). The upstream southward IMF $B$ field is in the $-z$ direction, and thus the main compressed magnetic field component in the shock transition is $B_z$. The Mach number of the incident solar wind along the shock normal is $V_{SW}\times sin\theta_{Bn}=8V_A$. Figure 3c shows the shock is in the supercritical regime and the front is undergoing self-reformation. For example, at about $t=10.8\Omega_{ci}^{-1}$, a foot (at $x=7d_i$) is formed ahead of the shock ramp ($x=8.5d_i$). The newborn foot propagates together with the injected solar wind toward the right hand side. Finally, the foot grows and becomes a new ramp at about $t=12.5\Omega_{ci}^{-1}$ and the process repeats itself. Different reformation cycles are marked by red arrows. The observed period of the reformation cycle is consistent with previous local hybrid and PIC simulations \citep{Lembege1992,Matsukiyo2003,Yang2009,Yuan2009}.

Figures 3d-3f display similar plots for the shock profile B, which is a quasi-perpendicular shock ($\theta_{Bn}\approx70^o$) measured at $z=66d_i$ (see Figure 2g). First, the magnetic field component $B_x$ measured in the southern part of the bow shock is almost positive (e.g., the profile A) and that measured in the northern part is negative (profiles B, C, and D). It depends on the sampling locations (refer Figure 2g) because the IMF is curved inside the magnetosheath. Second, the Rankine-Hugoniot equations specify only the change in magnetic field and plasma parameters from one side of the shock to the other and do not specify how these parameters change inside the shock. Thus, while the upstream, downstream magnetic field, and the shock normal $\hat{n}$ must all be in the same plane, the magnetic field within the shock layer could deviate from the plane and become noncoplanar. The magnetic field in fact frequently deviates from this plane both in observations of Earth's bow shock \citep{Friedman1990}, in numerical simulations of collisionless shocks \citep{Thomsen1987}, and theoretical models that including at least two fluids \citep{Gedalin1996}. In our simulation, the component $B_y$ is the non coplanar magnetic field component (Figure 3e). Third, the Mach number of the incident solar wind along the shock normal is $V_{SW}\times sin\theta_{Bn}=7.5V_A$ which is smaller than that in profile A, but the shock is still in the supercritical regime. Hence, the shock front is still nonstationary, and the self-reformation process takes place. However, the maximum amplitude of the profile $|B_x|$ at the overshoot is lower than the profile for A.

Figures 3g-3i and 3j-3l display the stack-plots for shock profiles B and C, respectively. These two shocks are sampled at the oblique shock region ($z=81d_i$ and $96d_i$) of the bow shock. Their corresponding shock normal angles $\theta_{Bn}$ are about $50^o$ and $30^o$, respectively. The Mach number of the incident solar wind along the shock normal for these two shocks are $V_{SW}\times sin\theta_{Bn}=6.1V_A$ and $4V_A$, respectively. Due to the decreased upstream solar wind speed in the shock normal direction, shock compression at the wings of the bow shock is weaker than that at the subsolar point as expected.

In contrast to the 1D shock model, the downstream region of a 2D bow shock is more complicated, because the incident and transmitted solar wind particles at the perpendicular region can convect to the downstream region of the quasi-perpendicular and oblique shocks. The bulk velocities of ions and electrons in the downstream have already been shown in Figures 2c and 2f. To show the single particle motions in the magnetosheath, we have traced the particle trajectories from the simulation. Figures 4a and 4b show the trajectories of solar wind ions and electrons injected at the nearly perpendicular shock (i.e., subsolar point). The locations of the bow shock (black curve) and the magnetosphere (black dashed) are also shown for reference. These locations are obtained by tracing their visual outlines shown in Figure 2. Initially, the ions and electrons drift together in the solar wind toward the bow shock (black dots). Later on, the incident particles reach the bow shock (blue dots). Then a fraction of the incident ions is reflected by the shock leading to a self-reformation of the nonstationary shock front, and almost all of the electrons are directly transmitted across the shock (green dots). At a later time, both ions and electrons diffuse into the downstream region and the particles are convected to the downstream regions of quasi-perpendicular and oblique shock regions (yellow dots). Finally, the particles become more dispersed and a fraction of electrons leak out from the magnetosheath at the wings of the bow shock, that is at the oblique shock region (red dots). Figures 4c and 4d display similar plots for ions and electrons with different initial locations (the particles injected at the northern part of the bow shock). A large number of ions and electrons can be reflected at the oblique shock front. At a later time, a fraction of ions can enter the magnetosphere on the nightside. Some electrons can be trapped by the dipole field and these trapped electrons include bounce motion on the magnetic field on the dayside. These particles could affect the ring current in the inner magnetosphere, but the analysis requires at least a 3D model which is beyond the scope of this article. Other electrons are convected to the southern part of the bow shock.

\subsection{Impact of $\beta$ and $M_A$ on the shock nonstationarity}

In this part, we study the impact of the plasma $\beta_i$ and the solar wind Mach number $M_A$ on the shock front nonstationarity. \textbf{It is generally thought that the quasi-perpendicular shock front self-reformation and local instabilities can be affected by the ion beta and the upstream inflow speed \citep{Hellinger2002,Scholar2004,Yang2013}.} The results of the baseline case (Run 1: $\beta_i=0.01$ and $M_A=8$) have already been shown in Sections 3.1 and 3.2 above. Two comparative cases have been carried out: (1) Run 2, $\beta_i=2$ and $M_A=8$, and (2) Run 3, $\beta_i=2$ and $M_A=5$. The other setups are kept unchanged. In run 2, we increase the ion beta value only. In run 3, we increase the ion beta but decrease the solar wind Mach. Figure 5 shows the main magnetic field component $B_z$ from Runs 1, 2, and 3, respectively. By comparing Figures 5a and 5b, we find that the ripples on the bow shock surface become blurred and the spatial scale of the rippling wave length becomes larger. However, the bow shock is still nonstationary. The magnetosheath at the subsolar point becomes thicker in Run 2. In addition, the magnetotail reconnection is visible in Runs 1 and 2. In contrast, Figure 5c shows the low Mach number case (Run 3). In this case, the shock front is quasi-stationary, and the magnetotail reconnection disappears. To double check the shock front nonstationarity in different runs, the main magnetic field component $B_z$ is sampled at different locations of the bow shock. Figures 6e-6h show the stack-plots of the shock profiles measured at different locations of the bow shock obtained in Run 3 at different times. The corresponding plots for Run 1 are also shown for reference in Figures 7a-7d. It should be clear that the bow shock with a high plasma beta value and a low solar wind speed is quasi-stationary. This behavior is in consistent with previous 1D and 2D local simulations of planar shocks \citep{Scholar2004,Yang2013}.

\section{Summary and Discussion}

In this paper, we have presented a 2D global PIC model to study the interaction between the solar wind and magnetosphere. We demonstrate the capability of the model focusing on the kinetic effects associated with the bow shock nonstationarity and magnetic reconnections. By sampling the shock profiles at different times and different locations of the bow shock and tracing the particles, we have shown the following results:

1. We identify different stages in the macroscopic evolution of the bow shock and the entire formation of the magnetosphere in the meridian plane. The macrostructures reach a quasi-stable state after several ion cyclotron cycles, consistent with previous 3D implicit PIC simulations \citep{Peng2015a}. Furthermore, a relatively high temporal-spatial resolution employed in our simulation provides an opportunity to examine the microstructure of the bow shock and the magnetic reconnection.

2. In the southward IMF condition, the shock around the subsolar point is quasi-perpendicular, and the shock at the wings is oblique. The angle $\theta_{Bn}$ and solar wind speed along the shock normal decrease with the distance away from the subsolar point. At quasi-perpendicular regions of the shock, a self-reformation of the shock front is found and the cyclic period of the reformation is similar to those observed in 1D and 2D local simulations. At the oblique region, the shock becomes weak due to the slow solar wind speed in the shock normal direction.

3. Different from 1D and 2D local simulations, particles in the downstream region are more complicated at the bow shock. By tracing the ions and electrons in the simulation, we find that the solar wind ions injected at the subsolar point will become diffuse and can be convected to the cusp region and the downstream of the oblique region of the bow shock. Electrons are more diffusive than ions in the downstream region. The ions injected at the northern part of the bow shock will fill in the northern part of the magnetosheath. A fraction of electrons injected at the same place will be convected to the southern part of the magnetosheath. The particles injected and reflected by the quasi-perpendicular shock region can affect the local velocity distributions and electromagnetic fluctuations in the transition and downstream regions of the oblique shock at the wings of the bow shock. This feature makes the oblique shock regions become a little different from those observed in local planar shock simulations.

4. The impact of the upstream plasma beta $\beta_i$ and solar wind Mach number $M_A$ on the bow shock and the magnetosphere show that the shock front becomes quasi-stationary in the high beta and low Mach number cases. In addition, the magnetotail reconnection disappears in this quasi-stationary case due to the low compression of the magnetotail under a low solar wind pressure condition. If we keep the Mach number of the incident solar wind unchanged and only increase the beta value, the bow shock is still nonstationary. Contrasting the low beta case, the ripple scale becomes larger in the high beta case.

5. To confirm the impact of the solar wind condition on the bow shock front nonstationarity, we also studied the time-evolution of shock profiles sampled at different locations of the bow shock. The results support the conclusions in point 4.

\textbf{It is worth noting that there are two limitations in PIC simulations: (1) the unrealistic mass ratio $m_i/m_e$ and (2) the unrealistic ratio of electron plasma frequency to cyclotron frequency $\omega_{pe}/\Omega_{ce}=\frac{c}{v_A}\sqrt{\frac{m_e}{m_i}}$. \citet{Quest1986} has point out that the use small values of $\omega_{pe}/\Omega_{ce}$ overemphasizes charge separation effects, and Krasnoselskikh et al. [2013] note that the electric field fluctuations are overestimated in PIC simulations with low values of $\omega_{pe}/\Omega_{ce}$. The impact of these two parameters on the shock front self-reformation has been investigated by 1D PIC simulations \citep{Matsukiyo2003}. They conclude that: (1) in small ion to electron mass ratio runs, the reformation is due to the accumulation of gyrating reflected ions. Furthermore, at the extremely small mass ratio, the Buneman instability is generated in the foot. In the realistic mass ratio run, however, the modified two-stream instability excited in the foot leads to the reformation. Hence, the self-reformation also occurs in the realistic mass ratio run but the associated instability is changed. Of course, a higher mass ratio allows for an easier separation between ion and electron scales but requires computer capacities relatively large, in particular for 2D or 3D simulations; (2) The self-reformation is not a low $\omega_{pe}/\Omega_{ce}$ process but occurs also in $(\omega_{pe}/\Omega_{ce})^{2}\gg1$. Nevertheless, they do mention that in the solar wind at the Earth¡¯s orbit the quantity of $\omega_{pe}/\Omega_{ce}$ is 100-200. However, in most simulations the value of $\omega_{pe}/\Omega_{ce}$ is assumed to be of the order of 1, i.e., simulations are done for shocks in the strongly magnetized condition because of computational constraints of PIC codes. In summary, recent 1D simulations have evidenced that the shock front self-reformation can occur even in high mass ratio and high plasma to cyclotron frequency ratio conditions. The impact of such parameters on self-reformation in 2D and 3D simulations is still open due to the enormous computation burden.}

Future work will include the background turbulence in the solar wind because hybrid simulations show that the turbulence can affect the bow shock and magnetosphere structures \citep{Guo2012,Karimabadi2014}. \textbf{Furthermore, it is interesting to see the impact of the third population of $O^+$ ion outflows \citep{Seki2001,Lennartsson2004,Wiltberger2010} on the magnetotail reconnection by using the global PIC code. This has implications for understanding how planets begin to experience a runway greenhouse effect \citep{Zhang2012}.}

\acknowledgments The authors are grateful to D. S. Cai, X. Y. Wang, Y. Lin and G. Lapenta for helpful discussions on numerical methods and global simulations. This research was supported by NSFC under Grants No. 41574140, 41374173, the Recruitment Program of Global Experts of China, the Specialized Research Fund for State Key Laboratories of China.


\begin{thebibliography}{}

\bibitem[Baraka \& Ben-Jaffel(2005)]{Baraka2007} \textbf{Baraka, S. \& Ben-Jaffel, L. 2007, J. Geophys. Res., 112, A06212.}

\bibitem[Buneman et al.(1995)]{Buneman1995} Buneman, O., Nishikawa, K.-I., \& Neubert, T. 1995, AGU Geophys. Monograph, 86, 347.

\bibitem[Burgess et al.(2005)]{Burgess2005} Burgess, D., Lucek, E. A., Scholer, M., et al. 2005, Space Sci. Rev., 110, 161.

\bibitem[Burgess \& Scholer(2007)]{Burgess2007} Burgess, D., \& Scholer, M. 2007, Phys. Plasmas, 14, 012108.

\bibitem[Burlaga et al.(2008)]{Burlaga2008} Burlaga, L. F., Ness, N. F., Acu\~{n}a, M. H., et al. Nature, 454, 07029.

\bibitem[Cai et al.(2015)]{Cai2015} Cai, D., Esmaeili, A., Lemb\`ege, B., \& Nishikawa, K.-I. 2015, J. Geophys. Res., 120, 8368.

\bibitem[Daughton et al.(2006)]{Daughton2006} Daughton, W., Scudder, J., \& Karimabadi, H. 2006, Phys. Plasmas, 13, 072101.

\bibitem[Farrugia et al.(1998)]{Farrugia1998} Farrugia, C. J., Gratton, F. T., Bender, L., et al. 1998, J. Geophys. Res., 103, 6703.

\bibitem[Fu et al.(2006)]{Fu2006} Fu, X. R., Lu, Q. M., \& Wang, S. 2006, Phys. Plasmas, 13, 012309.

\bibitem[Friedman et al.(1990)]{Friedman1990} Friedman, M. A., Russell, C. T., Gosling, J. T., et al. 1990, J. Geophys. Res., 95, A3, 2441.

\bibitem[Gedalin(1996)]{Gedalin1996} Gedalin, M. 1996, J. Geophys. Res., 101, A5, 11153.

\bibitem[Guo \& Giacalone(2012)]{Guo2012} Guo, F., \& Giacalone, J. 2012, ApJ, 753, 1.

\bibitem[Guo et al.(2015)]{Guo2015} Guo, F., Liu, Y.-H., Daughton, W., et al. 2015, ApJ, 806, 167.

\bibitem[Hada et al.(2003)]{Hada2003} Hada, T., Oonishi, M., Lemb\`ege, B., et al. 2003, J. Geophys. Res., 108, A6, 1223.

\bibitem[Hao et al.(2014)]{Hao2014} Hao, Y. F., Lu, Q. M., Gao, X. L., et al. 2014, J. Geophys. Res., 119, 3225.

\bibitem[Hellinger et al.(2002)]{Hellinger2002} \textbf{Hellinger, P., Tr\'{a}vn\'{\i}\v{c}ek, \& Matsumoto, H. 2002, Geophys. Res. Lett., 29, No. 24, 2234.}

\bibitem[Hellinger et al.(2007)]{Hellinger2007} Hellinger, P., Tr\'{a}vn\'{\i}\v{c}ek, P., Lemb\`ege, B., et al. 2007, Geophys. Res. Lett., 34, L14109.

\bibitem[Huang et al.(2014)]{Huang2014} Huang, C., Lu, Q. M., Lu, S., et al. 2014, J. Geophys. Res., 119, 798.

\bibitem[Karimabadi et al.(2014)]{Karimabadi2014} Karimabadi, H., Roytershteyn, V., Vu, H. X., et al. 2014, Phys. Plasmas, 21, 062308.

\bibitem[Kempf et al.(2015)]{Kempf2015} Kempf, Y., Pokhotelov, D., Gutynska, O., et al. 2015, J. Geophys. Res., 120, 3684.

\bibitem[Krasnoselskikh et al.(2002)]{Krasnoselskikh2002} Krasnoselskikh, V. V., Lemb\`ege, B., Savoini, P., \& Lobzin, V. V. 2002, Phys. Plasmas, 9, 1192.

\bibitem[Krasnoselskikh et al.(2013)]{Krasnoselskikh2013} \textbf{Krasnoselskikh, V. V., Balikhin, M., Walker, S. N., et al. 2013, Space Sci. Rev., 178, 535.}

\bibitem[Lee et al.(1996)]{Lee1996} Lee, M. A., Shapiro, V. D., \& Sagdeev, R. Z. 1996, J. Geophys. Res., 101, 4777.

\bibitem[Lee et al.(2004)]{Lee2004} Lee, R. E., Chapman, S. C., \& Dendy, R. O. 2004, ApJ, 604, 187.

\bibitem[Lemb\`ege \& Dawson(1987)]{Lembege1987} Lemb\`ege, B., \& Dawson, J. M. 1987, Phys. Fluids, 30, 1767.

\bibitem[Lemb\`ege \& Savoini(1992)]{Lembege1992} Lemb\`ege, B., \& Savoini, P. 1992, Phys. Fluids, 4, 3533.

\bibitem[Lemb\`ege et al.(2004)]{Lembege2004} Lemb\`ege, B., Giacalone, J., Scholer, M., et al. 2004, Space Sci. Rev., 118, 205.

\bibitem[Lemb\`ege et al.(2009)]{Lembege2009} Lemb\`ege, B., Savoini, P., Hellinger P., et al. 2009, J. Geophys. Res., 114, A03217.

\bibitem[Lennartsson et al.(2004)]{Lennartsson2004} Lennartsson, O. W., Collin, H. L., \& Peterson, W. K. 2004, J. Geophys. Res., 109, A12212.

\bibitem[Li et al.(2013)]{Li2013} Li, W. Y., Wang, C., Tang, B. B., et al. 2013, J. Geophys. Res., 118, 5118.

\bibitem[Lin(2003)]{Lin2003} Lin, Y. 2003, J. Geophys. Res., 108, 1390.

\bibitem[Lipatov \& Zank(1999)]{Lipatov1999} Lipatov, A. S., \& Zank, G. P. 1999, Phys. Rev. Lett., 82, 3609.

\bibitem[Lobzin et al.(2007)]{Lobzin2007} Lobzin, V. V., Krasnoselskikh, V. V., Pin\c{c}on, J.-L., et al. 2007, Geophys. Res. Lett., 34, L05107.

\bibitem[Lu et al.(2010)]{Lu2010} Lu, Q. M., Huang, C., Xie, J., et al. 2010, J. Geophys. Res., 115, A11208.

\bibitem[Lu et al.(2015)]{Lu2015} Lu, S., Lin, Y., Lu, Q. M., et al. 2015, Phys Plasmas, 22, 052901.

\bibitem[Matsukiyo et al.(2003)]{Matsukiyo2003} Matsukiyo, S., Shinohara, I., \& Scholer, M. 2003, J. Geophys. Res., 108, 1014.

\bibitem[Matsukiyo \& Scholer(2011)]{Matsukiyo2011} Matsukiyo, S., \& Scholer, M. 2011, J. Geophys. Res., 116, A08106.

\bibitem[Mazelle et al.(2010)]{Mazelle2010} Mazelle, C., Lemb\`ege, B., Morgenthaler, A., et al. 2010, AIP Conf.
    Proc., 1216, 471.

\bibitem[Moullard et al.(2006)]{Moullard2006} Moullard, O., Burgess, D., Horbury, T. S., et al. 2006, J. Geophys. Res., 111, A09113.

\bibitem[Ogino(1986)]{Ogino1986} Ogino, T. 1986, J. Geophys. Res., 91, 6791.

\bibitem[Omidi et al. (2005)]{Omidi2005} Omidi, N., Blanco-Cano, X., \& Russell, C. T. 2005, J. Geophys. Res., 110, A12212.

\bibitem[Omidi et al. (2006)]{Omidi2006} \textbf{Omidi, N., Blanco-Cano, X., Russell, C. T., \& Karimabadi, H. 2006, Adv. Space Res., 38, 632.}

\bibitem[Omidi \& Sibeck(2007)]{Omidi2007} Omidi, N., \& Sibeck, D. G. 2007, J. Geophys. Res., 112, A01203.

\bibitem[Omidi et al.(2010)]{Omidi2010} \textbf{Omidi, N., Eastwood, J. P., \& Sibeck, D. G. 2010, J. Geophys. Res., 115, A06204.}

\bibitem[Omidi et al.(2014)]{Omidi2014} \textbf{Omidi, N., Sibeck, D., Gutynska, O., and Trattner, K. J. 2014, J. Geophys. Res., 119, 2593.}

\bibitem[Palmroth et al.(2015)]{Palmroth2015} \textbf{Palmroth, M., Archer, M., Vainio, R., et al. 2015, J. Geophys. Res., 120, 6133.}

\bibitem[Peng et al.(2015a)]{Peng2015a} Peng, I. B., Markidis, S., Vaivads, A., et al. 2015, Proc. Comput. Sci., 51, 1178.

\bibitem[Peng et al.(2015b)]{Peng2015b} Peng, I. B., Markidis, S., Laure, E., et al. 2015, Phys. Plasmas, 22, 092109.

\bibitem[Quest(1985)]{Quest1985} Quest, K. B. 1985, Phys. Rev. Lett., 54, 1872.

\bibitem[Quest(1986)]{Quest1986} Quest, K. B. 1986, Los Alamos Natl. Lab. Rep. LAUR, 86.

\bibitem[Rekaa et al.(2014)]{Rekaa2014} Rekaa, V. L., Chapman, S. C., \& Dendy, R. O. ApJ, 791, 26.

\bibitem[Richardson et al.(2008)]{Richardson2008} Richardson, J. D., Kasper, J. C., Wang, C., et al. Nature, 454, 07024.

\bibitem[Richer et al.(2012)]{Richer2012} Richer, E. R., Modolo, R., Chanteur, G. M., et al. J. Geophys. Res., 117, A10228.

\bibitem[Savoini \& Lemb\`ege(1994)]{Savoini1994} \textbf{Savoini, P. \& Lemb\`ege, B. 1994, J. Geophys. Res., 99, A4, 6609.}

\bibitem[Scholer et al.(2003)]{Scholer2003} Scholer, M., Shinohara, I., \& Matsukiyo, S. 2003, J. Geophys. Res., 108, 1014.

\bibitem[Scholer \& Mastsukiyo(2004)]{Scholar2004} Scholer, M., \& Matsukiyo, S. 2004, Ann. Geophys., 22, 2345.

\bibitem[Seki et al.(2001)]{Seki2001} Seki, K., Elphic, R. C., Hirahara, M., et al. 2001, Science, 291, 1939.

\bibitem[Sundberg et al.(2013)]{Sundberg2013} Sundberg, T., Boardsen, S. A., Slavin, J. A., et al. 2013, J. Geophys. Res., 118, 6457.

\bibitem[Swift(1995)]{Swift1995} Swift, D. W. 1995, Geophys. Res. Lett., 22, 311.

\bibitem[Thomsen et al.(1987)]{Thomsen1987} Thomsen, M. F., Gosling, J. T., Bame, S. J., et al. 1987, J. Geophys. Res., 92, A3, 2305.

\bibitem[Tidmann \& Krall(1971)]{Tidmann1971} Tidmann, D. A., \& Krall, N. A. 1971 (New york: Whiley).

\bibitem[Tiu et al.(2011)]{Tiu2011} Tiu, D., Cairns, I. H., Yuan, X. Q., et al. 2011, J. Geophys. Res., 116, A04228.

\bibitem[Tr\'{a}vn\'{\i}\v{c}ek et al.(2005)]{Travnicek2005} Tr\'{a}vn\'{\i}\v{c}ek, P., Hellinger, P., Schriver D., \& Bale, S. D. 2005, Geophys. Res. Lett., 32, L06102.

\bibitem[Tr\'{a}vn\'{\i}\v{c}ek et al.(2007)]{Travnicek2007} Tr\'{a}vn\'{\i}\v{c}ek, P., Hellinger, P., \& Schriver D. 2007, Geophys. Res. Lett., 34, L05104.

\bibitem[Turc et al.(2015)]{Turc2015} \textbf{Turc, L., Fontaine, D., Savoini, P., \& Modolo, R. 2015, J. Geophys. Res., 120, 6133.}

\bibitem[Umeda et al.(2008)]{Umeda2008} Umeda, T., Yamao, M., \& Yamazaki, R. 2008, ApJ, 681, L85.

\bibitem[Vapirev et al.(2013)]{Vapirev2013} Vapirev, A. E., Lapenta, G., Divin, A., et al. 2014, J. Geophys. Res., 118, 1435.

\bibitem[Wang et al.(2014)]{Wang2014} Wang, C., Han, J. P., Li H., et al. 2014, J. Geophys. Res., 119, 6199.

\bibitem[Wiltberger et al.(2010)]{Wiltberger2010} Wiltberger, M., Lotko, W., Lyon, J. G., et al. 2010, J. Geophys. Res., 115, A00J05.

\bibitem[Winske et al.(1990)]{Winske1990} Winske, D., Omidi, N., Quest, K. B., et al. 1990, Hybrid Simulation Codes: Past, Present and Future-A Tutorial, e.d. B\"{u}chner, J., Dum, C. T., \& Scholer, M. (Verlag Berlin Heidelberg: Springer), 136.

\bibitem[Winske et al.(2003)]{Winske2003} Winske, D., Lin, Y., Omidi, N., et al. 2003 (Germany: Springer).

\bibitem[Yang et al.(2009)]{Yang2009} Yang, Z. W., Lu, Q. M., Lemb\`ege, et al. 2009, J. Geophys. Res., 114, A03111.

\bibitem[Yang et al.(2013)]{Yang2013} Yang, Z. W., Lu, Q. M., Gao, X. L., et al. 2013, Phys. Plasmas, 20, 092116.

\bibitem[Yang et al.(2015)]{Yang2015} Yang, Z. W., Liu, Y. D., Richardson, J. D., et al. 2015, ApJ, 809, 1.

\bibitem[Yuan et al.(2009)]{Yuan2009} Yuan, X., Cairns, I. H., Trichtchenko, L., et al. 2009, Geophys. Rev. Lett., 36, L05103.

\bibitem[Zank et al.(1996)]{Zank1996} Zank, G. P., Pauls, H. L., Cairns, I. H., et al. 1996, J. Geophys. Res., 101, 457.

\bibitem[Zhang et al.(2012)]{Zhang2012} Zhang, T. L., Lu, Q. M., Baumjohann, W., et al. 2012, Science, 336, 567.

\end{thebibliography}

\clearpage

\begin{sidewaystable}[t]
\small
\renewcommand{\arraystretch}{0.95}
\caption{Summary of simulation setups of hybrid and PIC global simulations of the magnetosphere)}
\centering
\label{table:1}
\begin{tabular}{@{}lccccccccc@{}}
\tableline
Simulations                     &  \textbf{Particles for each species per cell}        &spatial resolution $\Delta x$           & temporal resolution $\Delta t$\\
\tableline
 1. Global hybrid simulations \\
 \citet{Omidi2005}               & \textbf{- }                   & 0.5$d_i$                    & 0.0025$\Omega_{ci}^{-1}$\\
 \textbf{\citet{Omidi2010}}      & \textbf{15 }                  & \textbf{1$d_i$}                      & \textbf{-}\\
 \citet{Travnicek2005}           & \textbf{- }                   & 0.2-0.25$d_i$               & 0.005$\Omega_{ci}^{-1}$\\
 \citet{Richer2012}              & \textbf{- }                   & 3$d_i$                      & 0.05$\Omega_{ci}^{-1}$\\
 \textbf{\citet{Lu2015}}          & \textbf{50 (private communication)}  & \textbf{2$d_i$}               & \textbf{0.05$\Omega_{ci}^{-1}$}\\
 \tableline
 2. Global PIC simulations \\
 \textbf{\citet{Baraka2007} (TRISTAN) }             & \textbf{0.8 }              & 10$\lambda_{De}$        & -\\
 \citet{Cai2015} (TRISTAN)                 & \textbf{8 }                & 0.1$d_i$                & 0.005$\Omega_{ci}^{-1}$\\
 \citet{Vapirev2013} (iPic3D)              & \textbf{250}               & 0.078$d_i$              & 0.125$\Omega_{ci}^{-1}$\\
 \citet{Peng2015a,Peng2015b} (iPic3D)      & \textbf{25}                & 0.05-0.12$d_i$          & 0.1-0.15$\Omega_{ci}^{-1}$\\
 \tableline
 3. Local PIC simulations (Reforming shocks)\\
 \textbf{\citet{Savoini1994} (2D)}             & \textbf{4 }              & \textbf{0.79$\lambda_{De}$}         & \textbf{0.0012$\Omega_{ci}^{-1}$}\\
 \citet{Lee2004} (1D)                 & \textbf{200 }              & 1$\lambda_{De}$            & -\\
 \citet{Lembege2009} (2D)             & \textbf{4 }                & 0.0167$d_i$                & 7.5$\times10^{-5}\Omega_{ci}^{-1}$\\
 \citet{Matsukiyo2011} (1D)           & \textbf{100}               & 0.89$\lambda_{De}$         & 1.2$\times10^{-5}\Omega_{ci}^{-1}$\\
 \tableline
 4. \textbf{This work}                  & \textbf{4}                 & 0.025$d_i\ (0.68\lambda_{De})$               & 0.001$\Omega_{ci}^{-1}$\\
\tableline
\end{tabular}
\end{sidewaystable}

\clearpage

\begin{figure}
\epsscale{0.8} \plotone{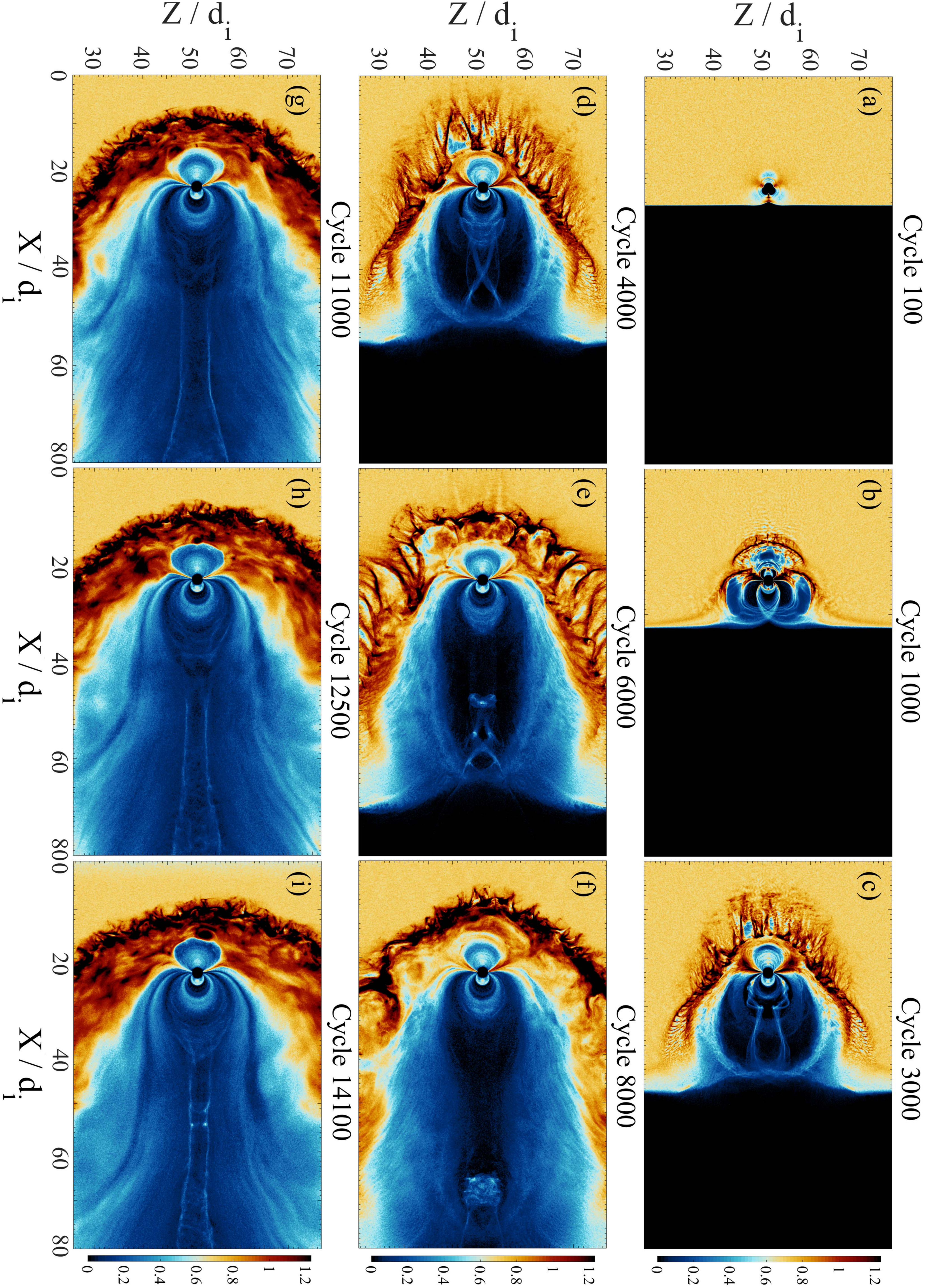}
\caption{Evolution of the ion density in the meridian plane ($x-z$ plane). \textbf{The color bar indicates the value of $log_{10}(N_i+1)$, where $N_i$ (equals to 4 in the upstream region) is the count number of the ions at each grid. The ion count number $N_i$ for color values ranges from 0 to $\gtrsim$16.} (a-c): The magnetosheath expands in the initial states, when the magnetotail stretches along the $x$ direction. (d-f): The bow shock forms, while the magnetosheath plasmas are compressed. (g-i): The bow shock reaches a steady state after about 10000 cycles with minimal changes till the end of the simulation. The color bar is log of ion density.
\protect\\}
\end{figure}

\clearpage

\begin{figure}
\epsscale{0.95} \plotone{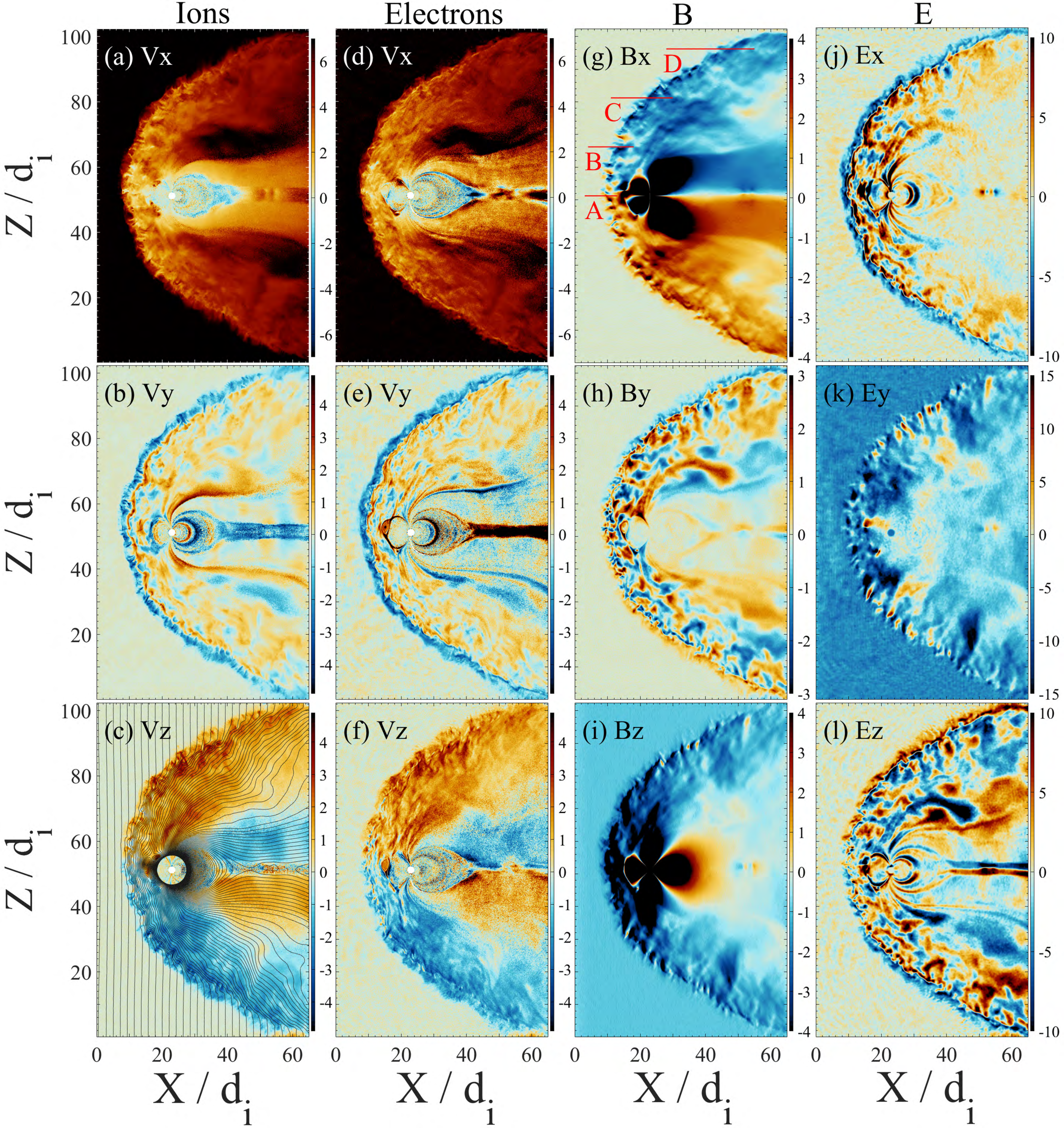}
\caption{(a-c): Contours of ion bulk velocity components at $t=14\Omega_{ci}^{-1}$ along $x$, $y$, and $z$ directions, respectively. (d-f): Corresponding velocity components of the electrons. (g-i): Magnetic field components. (j)-(l): Electric field components.
\protect\\}
\end{figure}

\clearpage

\begin{figure}
\epsscale{0.98} \plotone{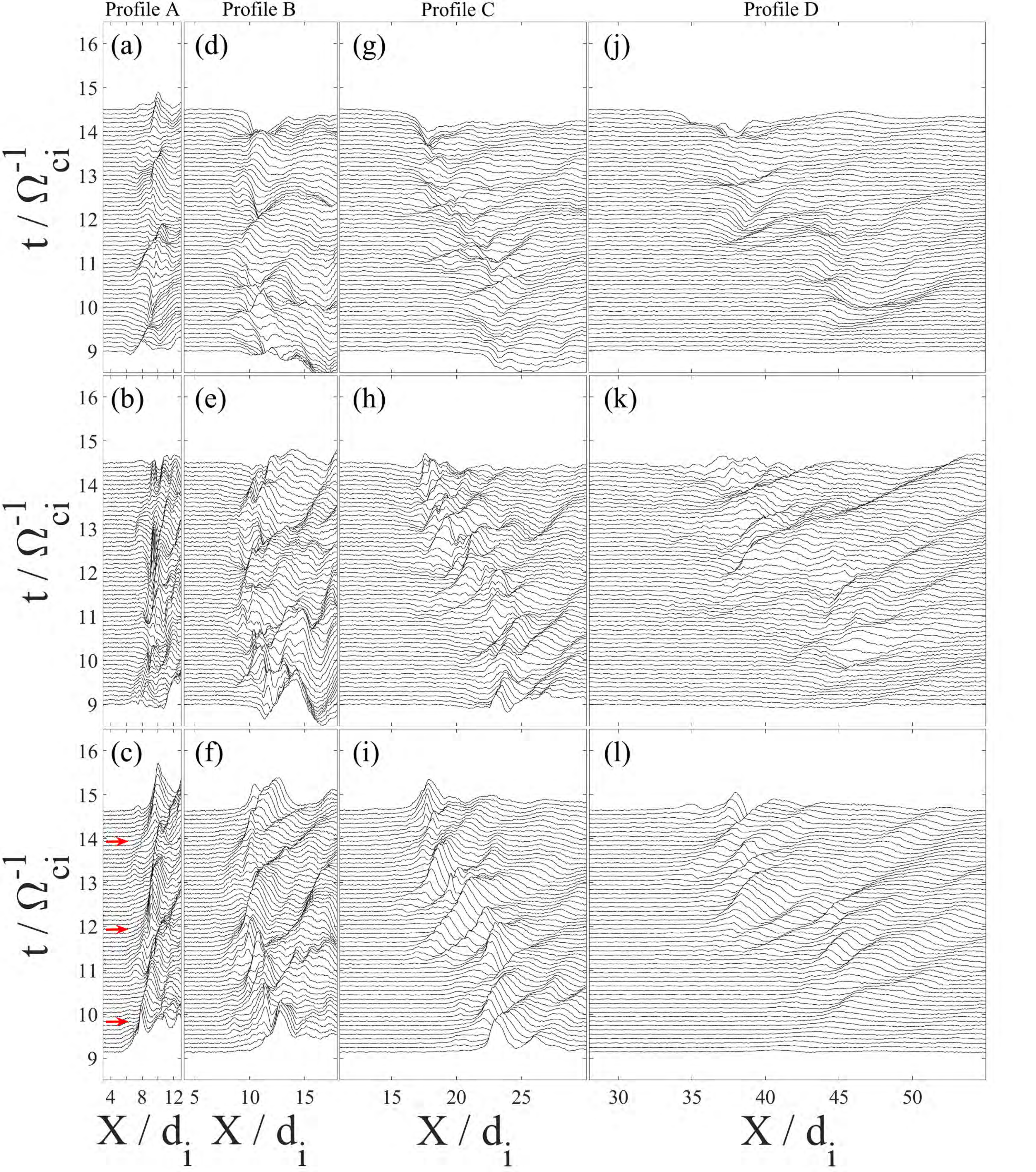}
\caption{Time evolution (stack-plots) of the shock magnetic field components $B_x$ (top), $B_y$ (middle), and $-B_z$ (bottom) measured from the perpendicular region to the oblique region at the bow shock. The profiles A, B, C, and D measured at $z=51.2d_i$ (1st column), $66.2d_i$ (2nd column), $81.2d_i$ (3rd column), and $96.2d_i$ (4th column), respectively. The spatial ranges of measurements are marked by the red lines in Figure 2g. The red arrows in panel (c) indicate the self-reformation of the shock front in the shock rest frame.
\protect\\}
\end{figure}

\clearpage

\begin{figure}
\epsscale{0.8} \plotone{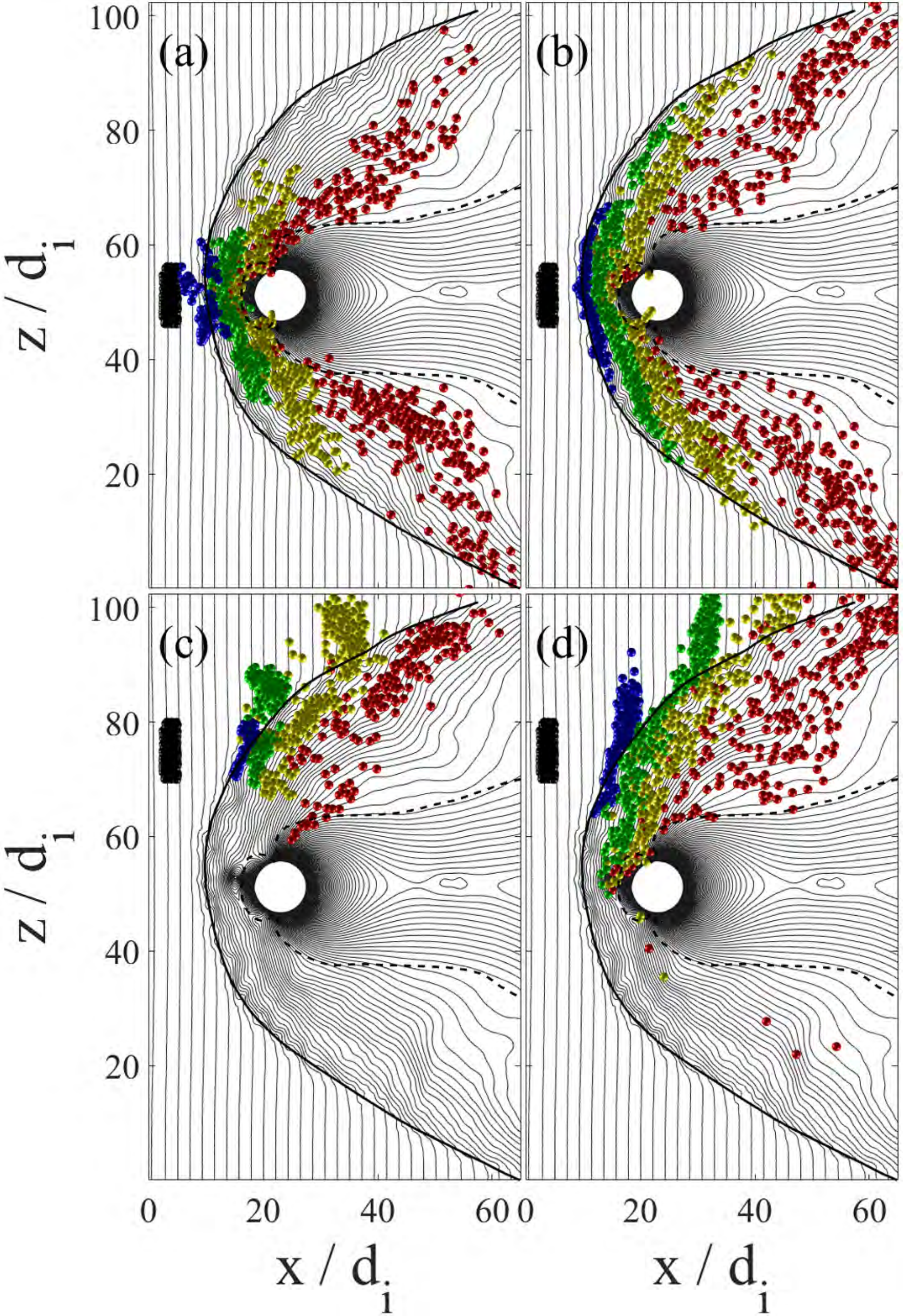}
\caption{Ion and electron trajectories. Top panels show the trajectories of ions (a) and electrons (b) injected at the subsolar point of the bow shock. Bottom panels show similar plots for ions (c) and electrons (d) injected in the northern part of the bow shock. Dots in different colors: black, blue, green, yellow, and red that indicate snapshots of particles in the time order. The black solid and dashed curves represent the locations of the bow shock and the magnetopause, respectively. The magnetic field lines are also shown for reference in thin gray curves.
\protect\\}
\end{figure}

\clearpage

\begin{figure}
\epsscale{0.95} \plotone{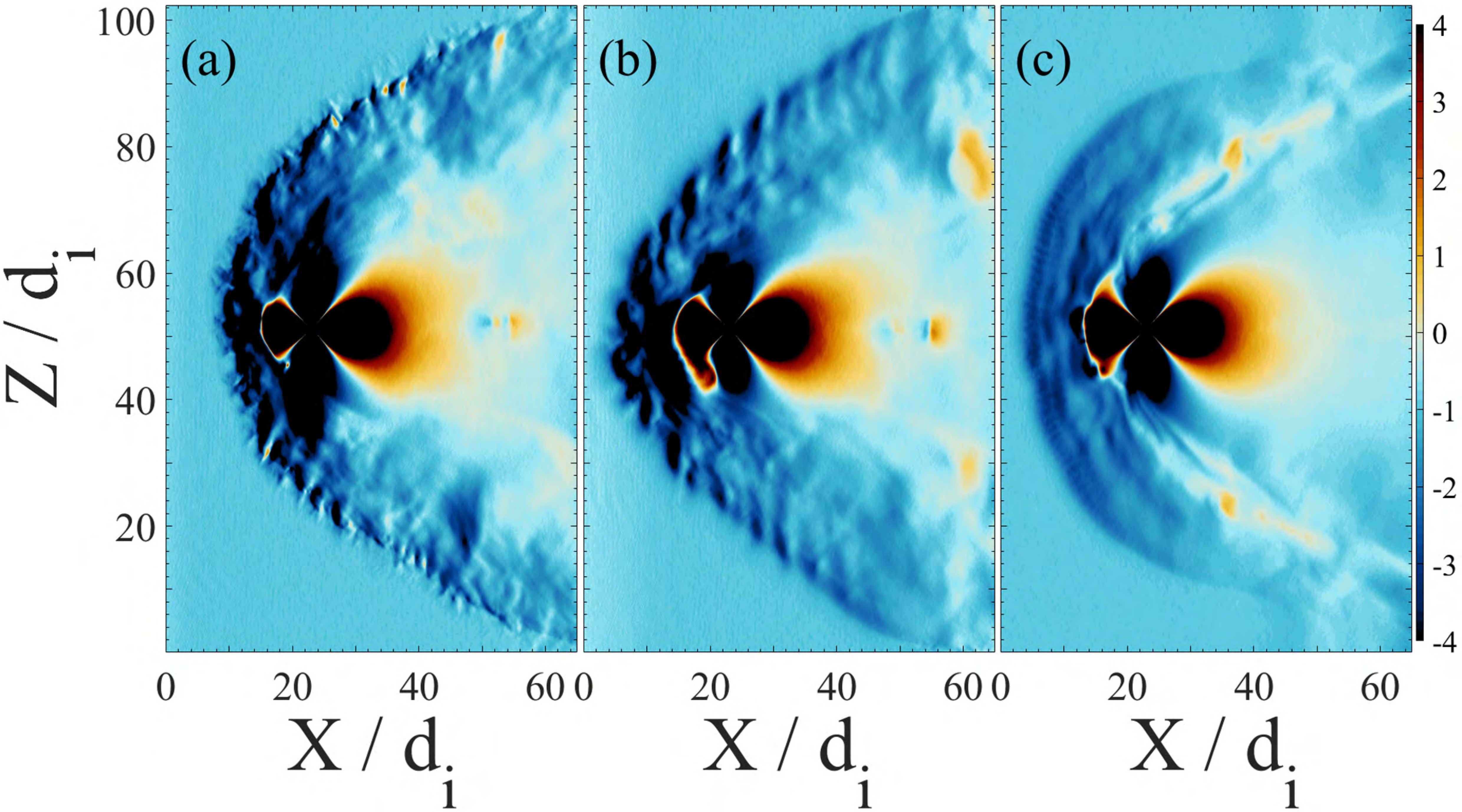}
\caption{Main magnetic field component $B_z$ at $t=14\Omega_{ci}^{-1}$ in Run 1 (a), Run 2 (b), and Run 3 (c). The color bar is in the same scale.
\protect\\}
\end{figure}

\clearpage

\begin{figure}
\epsscale{0.95} \plotone{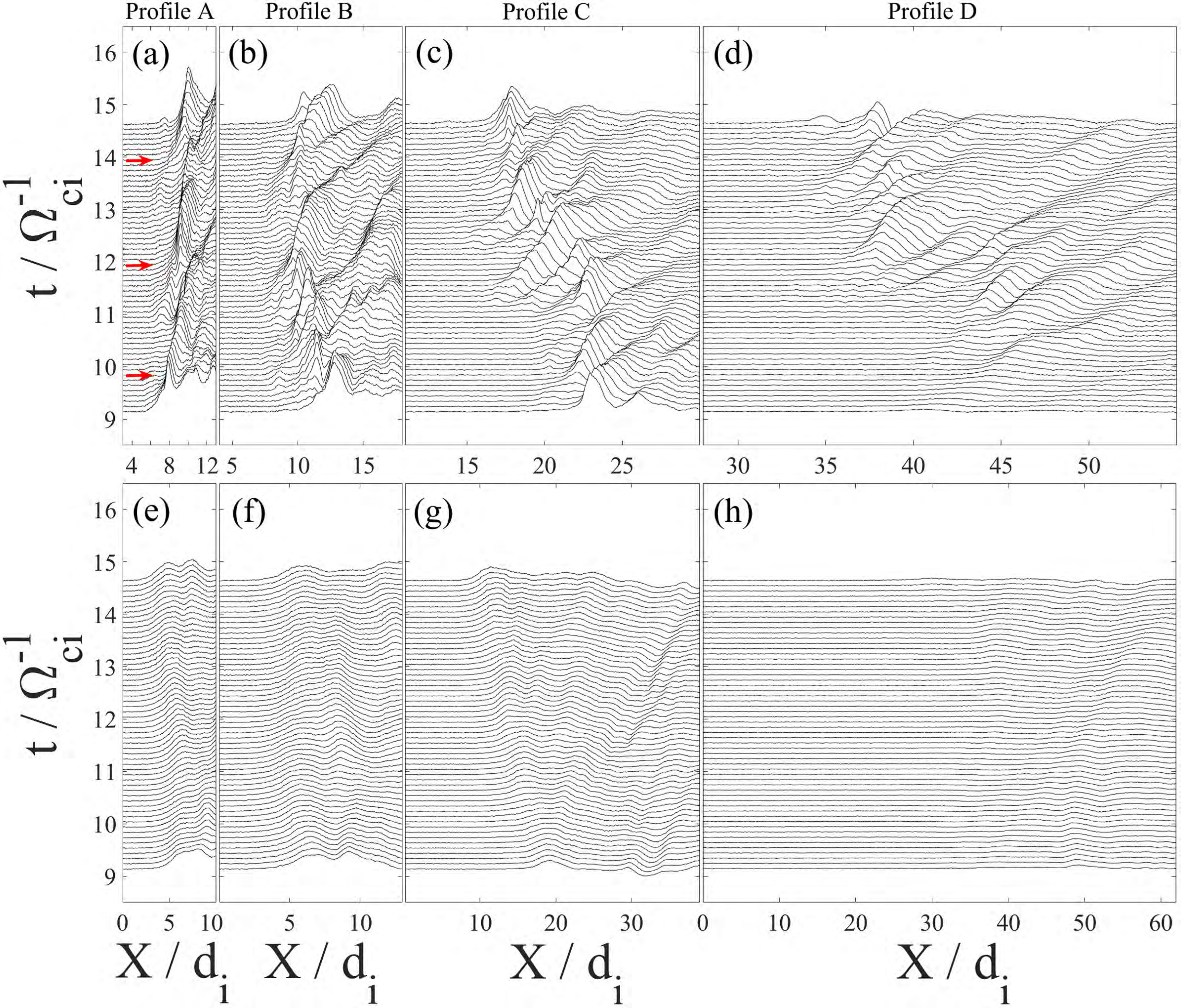}
\caption{Comparison of the shock front nonstationarity between Run 1 (a-d) and Run 3 (e-h). The sampling method and sampling location of profile A, B, C, and D are the same as in Figure 3.
\protect\\}
\end{figure}

\clearpage

\end{document}